\documentclass[submission, Proceedings]{SciPost}

\hypersetup{
    colorlinks,
    linkcolor={red!50!black},
    citecolor={blue!50!black},
    urlcolor={blue!80!black}
}

\usepackage[bitstream-charter]{mathdesign}
\usepackage{tikz}
\usepackage[compat=1.1.0]{tikz-feynman}
\usepackage{color}
\tikzstyle{every picture}+=[remember picture]
\DeclareSymbolFont{usualmathcal}{OMS}{cmsy}{m}{n}
\DeclareSymbolFontAlphabet{\mathcal}{usualmathcal}

\begin{document}

\begin{center}{\Large \textbf{
Towards the computation of inclusive decay rates using lattice QCD
}}\end{center}

\begin{center}
  Paolo Gambino\textsuperscript{1,2},
  Shoji Hashimoto\textsuperscript{3,4},
  Sandro M\"achler\textsuperscript{1,5},
  Marco Panero\textsuperscript{1},
  Francesco Sanfilippo\textsuperscript{6},
  Silvano Simula\textsuperscript{6},
  Antonio Smecca\textsuperscript{1$\star$} and
  Nazario Tantalo\textsuperscript{7}
\end{center}

\begin{center}
  {\bf 1} Dipartimento di Fisica, Universit\`a di Torino \& INFN, Sezione di Torino,\\Via Pietro Giuria 1, I-10125 Turin, Italy
  \\
    {\bf 2} Max Planck Institute for Physics, F\"ohringer Ring 6, 80805 M\"unchen, Germany
  \\
    {\bf 3} Theory Center, Institute of Particle and Nuclear Studies, High Energy Accelerator Research Organization (KEK), Tsukuba 305-0801, Japan
 \\
   {\bf 4} School of High Energy Accelerator Science, The Graduate University for Advanced Studies (SOKENDAI), Tsukuba 305-0801, Japan
 \\
   {\bf 5} Physikinstitut, Universit\"at Z\"urich, Winterthurerstrasse 190, CH-8057 Z\"urich, Switzerland
 \\
   {\bf 6} INFN, Sezione di Roma Tre, Via della Vasca Navale 84, I-00146 Rome, Italy
 \\
   {\bf 7} Dipartimento di Fisica, Universit\`a di Roma ``Tor Vergata'' \& INFN, Sezione di Roma ``Tor Vergata'', Via della Ricerca Scientifica 1, I-00133 Rome, Italy
  \\
* antonio.smecca@unito.it
\end{center}

\begin{center}
\today
\end{center}


\definecolor{palegray}{gray}{0.95}
\begin{center}
\colorbox{palegray}{
  \begin{tabular}{rr}
  \begin{minipage}{0.1\textwidth}
    \includegraphics[width=23mm]{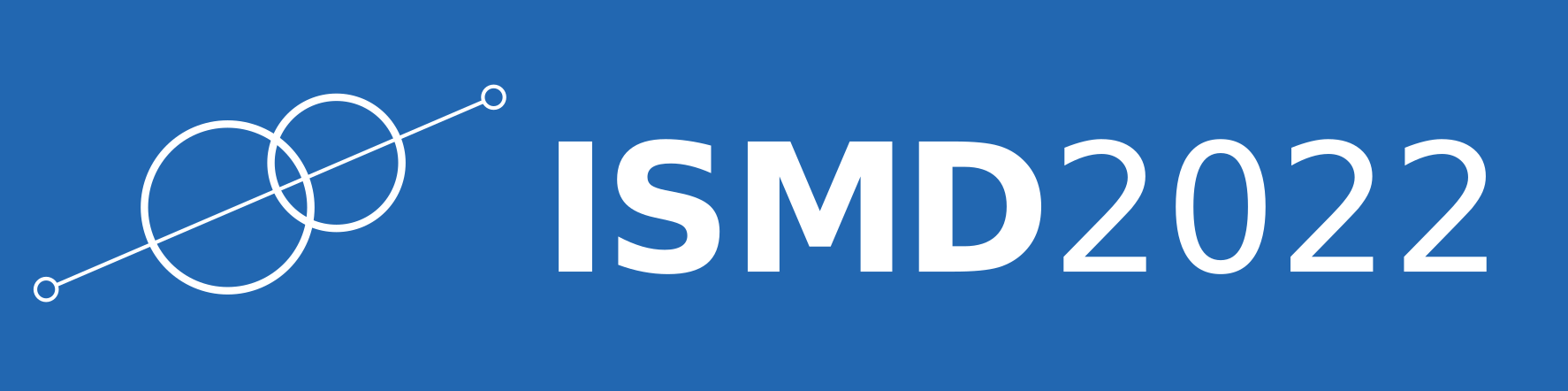}
  \end{minipage}
  &
  \begin{minipage}{0.8\textwidth}
    \begin{center}
    {\it 51st International Symposium on Multiparticle Dynamics (ISMD2022)}\\ 
    {\it Pitlochry, Scottish Highlands, 1-5 August 2022} \\
    \doi{10.21468/SciPostPhysProc.?}\\
    \end{center}
  \end{minipage}
\end{tabular}
}
\end{center}

\section*{Abstract}
{\bf
  We present a non-perturbative computation of inclusive rates of semileptonic decays of heavy mesons from lattice QCD simulations. The calculation is based on the extraction of smeared spectral functions obtained from four-point Euclidean correlation functions computed on configuration ensembles of the JLQCD and ETM collaborations. We compare our results for the inclusive decay rates with analytical predictions from the operator-product expansion, finding a good agreement for the calculation of the inclusive decay rate. This opens the path to the theoretical determination of the magnitude of the Cabibbo-Kobayashi-Maskawa matrix element $V_{cb}$ to a level of precision competitive with the present experimental uncertainty.
}


\section{Introduction}
\label{sec:intro}
Flavour physics remains an important avenue to access new physics that might exist beyond the Standard Model (SM), and in recent years the growing number of anomalies and tensions between SM predictions and experimental results in this sector has generated mounting excitement in the particle-physics community. One of the longest-standing tensions is the $\simeq 3 \sigma$ discrepancy between exclusive and inclusive determinations of the magnitude of $V_{cb}$, the Cabibbo-Kobayashi-Maskawa (CKM) matrix element~\cite{ParticleDataGroup:2020ssz, HFLAV:2019otj, Gambino:2019sif,Gambino:2020jvv}.
The exclusive determination of $|V_{cb}|$ requires non-perturbative form factors, which are typically computed in lattice QCD (LQCD), which has been very successful in providing precise and systematically improvable calculations~\cite{FlavourLatticeAveragingGroup:2019iem}.
Until recently, the LQCD approach was restricted to the study of exclusive semileptonic decays, while theoretical predictions for inclusive decays relied on the operator product expansion (OPE)~\cite{Wilson:1969zs, Kadanoff:1969zz}. In the past few years, however, new methods have been proposed, allowing a complete first-principle study of inclusive semileptonic decays using LQCD~\cite{Hashimoto:2017wqo,Gambino:2020crt}.

In this contribution, we discuss our recent lattice QCD study of inclusive semileptonic decays of $B_s$ mesons~\cite{Gambino:2022dvu}. In particular, the main focus of the analysis carried out in ref.~\cite{Gambino:2022dvu} consists in the application of the spectral function reconstruction to the four-point correlators extracted from the lattice. In the analysis, the method proposed in~\cite{Hansen:2019idp} is applied for the first time to the computation of inclusive decay rates and related observables.
We also compare the LQCD results with those obtained from the OPE, thereby testing the quark-hadron duality upon which the analytical approach is based.
\section{Theoretical framework}

Our determination of inclusive decay rates is based on the framework proposed in ref.~\cite{Gambino:2020crt}. The differential decay rate of the inclusive semileptonic decay of a $B_s$ meson to charmed final states $X_c$ and a pair of massless leptons $l\bar{\nu}$ is defined as
\begin{align}
  \frac{d\Gamma}{dq^2dq^0dE_\ell}=
  \frac{G_F^2|V_{cb}|^2}{8\pi^3} L_{\mu\nu}W^{\mu\nu},
  \label{eq:differential}
\end{align}
where $L_{\mu\nu}$ is the leptonic tensor and $W_{\mu\nu}$ is the hadronic tensor. The latter can be decomposed into Lorentz-invariant structure functions $Z^{(l)}(\omega,\boldsymbol{q}^2)$, and after integrating over the lepton energy $E_l$ the differential decay rate can be rewritten as
%
%
%
%
%
\begin{align}
  \frac{d\Gamma}{dq^2}=\frac{G_F^2|V_{cb}|^2}{24\pi^3|\boldsymbol{q}|}\sum_{l=0}^2 (\sqrt{q^2})^{2-l}Z^{(l)}(\boldsymbol{q}^2),
  \label{eq:IncDecay}
\end{align}
where $\omega$ is the energy of the final meson state in the rest frame of the $B_s$ meson and $Z^{(l)}(\boldsymbol{q}^2)$ is the integral over phase space of the structure functions $Z^{(l)}(\omega,\boldsymbol{q}^2)$ regulated by the integration kernel $\Theta^{l}(\omega_{max}-\omega)$:
\begin{align}
  Z^{(l)}(\boldsymbol{q}^2) = \int_0^\infty d\omega  \Theta^l(\omega_{\max}-\omega)Z^{(l)}(\omega,\boldsymbol{q}^2).
  \label{eq:Zfunction}
\end{align}
The integration kernel $\Theta^l$ is defined in terms of a step-function $\theta(\omega_{max}-\omega)$ and a kinematical factor: $\Theta^l=(\omega_{max}-\omega)^l \theta(\omega_{max}-\omega)$. Equation~(\ref{eq:Zfunction}) is one of the key equations of this work since it is essential in computing the inclusive differential decay rate defined in eq.~(\ref{eq:IncDecay}). The main goal of this work is to compute this quantity from lattice QCD simulations.

\section{Lattice computation}
The first step to compute the inclusive differential decay rate consists in evaluating the appropriate Euclidean correlation functions on the lattice. While lattice QCD calculations accessing only the ground state of a particle require only two- and three-point correlators, in ref.~\cite{Gambino:2020crt} it was shown that four-point functions allow one to access the full spectrum of charmed final states. The explicit form of the four-point correlation function in Euclidean time, with two electroweak currents sandwiched between $B_s$ meson states, is
\begin{align}
  C_{\mu\nu}(t_{\mathrm{snk}},t_2,t_1,t_{\mathrm{src}};\boldsymbol{q}) =        
  \int d^3x\, e^{i\boldsymbol{q}\cdot\boldsymbol{x}}\, 
  T\langle 0\vert\, \tilde{\phi}_B(\boldsymbol{0};t_{\mathrm{snk}})
  J_\mu^\dagger(\boldsymbol{x};t_2) J_\nu(\boldsymbol{0};t_1)
  \tilde{\phi}_B^\dagger(\boldsymbol{0};t_{\mathrm{src}})\, \vert 0\rangle .
\end{align}
In our analysis we focus on the contribution to the correlator between $t_1$ and $t_2$, in which multi-particle states propagate, and 
write a linear combination of the Euclidean correlator in terms of the associated hadronic-tensor components as
\begin{align}
  G^{(l)}(a\tau;\boldsymbol{q}) = \int_0^{\infty} d\omega Z^{(l)}(\omega,\boldsymbol{q}^2)e^{-\omega a\tau} ,
  \label{eq:Inverse_problem}
\end{align}
where we expressed the Euclidean time in units of the lattice spacing (denoted by $a$) as $a\tau$.

The two ensembles of lattice gauge-field configurations that we used in this work, generated by the JLQCD collaboration~\cite{Nakayama:2017lav, Colquhoun:2022atw} and by the ETM collaboration~\cite{Baron:2010bv, ETM:2010cqp, Frezzotti:2000nk, Frezzotti:2003xj, Frezzotti:2003ni, EuropeanTwistedMass:2014osg} rely on different types of discretization for the fermionic fields, but neither of them can accommodate a relativistic $b$ quark with its physical mass. As a consequence, our results are obtained with a $B_s$ meson that is lighter than in nature. For further details about the ensembles, see ref.~\cite{Gambino:2022dvu} and the references therein.

\section{Kernel reconstruction}
To relate the correlation functions extracted from the lattice with the differential decay rate in eq.~(\ref{eq:Zfunction}), one has to extract the associated spectral functions. This is an ill-posed inverse problem, which has to be tackled with appropriate numerical methods. 
In this work we use two methods, one based on Chebyshev polynomials~\cite{Bailas:2020qmv}, which we applied to the data obtained from the JLQCD ensemble, and another one based on the variant of the Backus-Gilbert method proposed in ref.~\cite{Hansen:2019idp}, which we used for the correlators extracted from the ETM configurations.
The starting point of both methods is that any smooth function $f(\omega)$ can be approximated numerically by a series of polynomials
\begin{align}
  f(\omega) = \sum_{\tau}^{\infty}g_{\tau}e^{-a\omega\tau},
\end{align}
so that eq.~(\ref{eq:Inverse_problem}) can be rewritten as
\begin{align}
  \sum_{\tau}^{\infty}g_{\tau}G^{(l)}(a\tau;\boldsymbol{q}) = \int_0^{\infty} d\omega Z^{(l)}(\omega;\boldsymbol{q}^2)f(\omega)\;.
\end{align}
The term on the r.h.s. of this equation is analogous to the expression in eq.~(\ref{eq:Zfunction}), provided one substitutes the kernel $\Theta^l(\omega_{max}-\omega)$ with $f(\omega)$. This substitution cannot be done straightforwardly, since the kernel $\Theta^l(\omega_{max}-\omega)$ is defined in terms of a step-function and therefore is not a smooth function. However, the substitution can be done with a smeared version of the step-function, $\theta_{\sigma}$, which yields a smooth kernel $\Theta^l_{\sigma}(\omega_{max}-\omega)$ that can then be written in terms of polynomials
\begin{align}
  \Theta^l_{\sigma}(\omega_{max}-\omega)
  = (\omega_{max}-\omega)^l\theta_{\sigma}(\omega_{max}-\omega)
  = m_{B_s}^l\sum_{\tau}^{\infty}
  g_{\tau}(\omega_{max};\sigma)e^{-a\omega\tau}\;,
\end{align}
which finally enables one to write eq.~(\ref{eq:Zfunction}) in terms of products of Euclidean correlation functions extracted from lattice QCD and coefficients of the smeared kernel:
\begin{align}
  Z_{\sigma}^{(l)}(\boldsymbol{q}^2) = \sum_{\tau}^{\infty}
  g_{\tau}(\omega_{max};\sigma) G^{(l)}(a\tau;\boldsymbol{q})\;.
  \label{eq:Z_function_G}
\end{align}
It is important to remark that eq.~(\ref{eq:Z_function_G}) describes a quantity defined in terms of a smeared kernel, which cannot be used to compute the physical decay rate. To extract the physical decay rate, one has to take the $\sigma \rightarrow 0$ extrapolation at the end of the calculations. The smearing procedure is not simply a numerical trick, it also has a meaning at the formal level. Indeed, hadronic spectral densities, and therefore also the structure functions $Z^{(l)}(\omega,\boldsymbol{q}^2)$, when they are computed in a finite volume such as in lattice simulations, have a discrete energy spectrum. Hence, to make the correct connection between $Z^{(l)}_{\sigma}$ and the corresponding physical quantity one has to take the infinite-volume limit \textit{first} and only then the $\sigma \rightarrow 0$ limit:
\begin{align}
  Z^{(l)}(\boldsymbol{q}^2)
  &= \lim_{\sigma \rightarrow 0}\left(\lim_{V \rightarrow \infty}\right)\int_0^{\infty}d\omega Z^{(l)}(\omega,\boldsymbol{q}^2)\Theta_{\sigma}^{l}(\omega_{max}-\omega) \nonumber \\
  &=  \lim_{\sigma \rightarrow 0}\left(\lim_{V \rightarrow \infty}\right)m_{B_s}^l\sum_{\tau}^{\infty}g_{\tau}G^{(l)}(a\tau;\boldsymbol{q})\;.
\end{align}
However, due to the exploratory nature of this work, we restricted our analysis only to one physical volume for each configuration ensemble, so that a $V \rightarrow \infty$ extrapolation was not possible. This choice can be justified by the fact that our present statistical uncertainties are likely larger than finite-volume effects; we plan to investigate these effects more thoroughly in future work with simulations in multiple volumes.


\section{Comparison with OPE results}
The quantity $Z^{(l)}(\boldsymbol{q})$ computed from lattice QCD simulations can be used to determine the differential decay rate according to eq.~(\ref{eq:IncDecay}). As the masses of the $B_s$ meson in the lattice QCD simulations differ slightly between the JLQCD and ETM ensembles, the final results cannot be compared directly with each other; however, both of them can be compared with the analytic predictions from the OPE.
\begin{figure}[h]
  \centering
  \includegraphics[width=10cm]{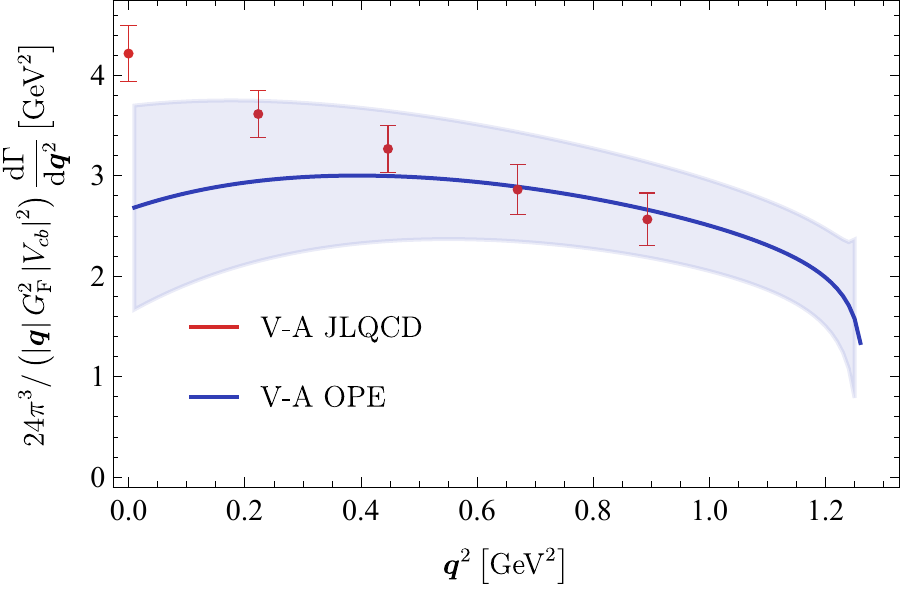}
  \includegraphics[width=10cm]{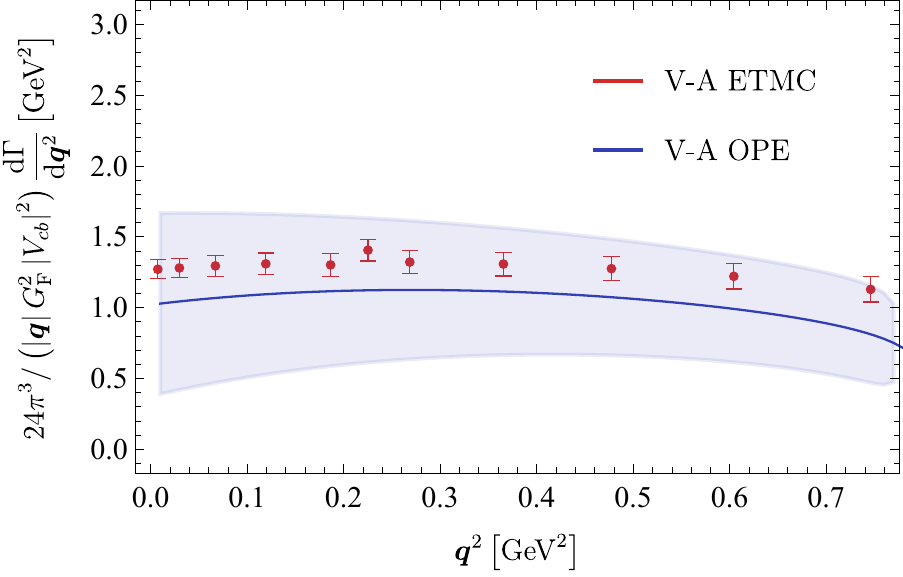}  
\caption{Differential $\boldsymbol{q}^2$ spectrum, divided by $|\boldsymbol{q}|$, in the SM. Comparison of OPE predictions with lattice QCD results from the JLQCD (top panel) and ETM (bottom panel) ensembles are shown.}
\label{fig:Comp}
\end{figure}
As shown in figure~\ref{fig:Comp}, the agreement between the OPE curve (in blue) and the lattice data (in red) is very good, for both the JLQCD and the ETM ensembles.

One can then perform the integration over $\boldsymbol{q}^2$ to obtain the inclusive decay rate for the $B_s \rightarrow X_cl\nu$ semileptonic decay. The final lattice result obtained using the JLQCD ensemble is $\frac{\Gamma}{|V_{cb}|^2}\times10^{13} = 4.46(21)$~GeV; the corresponding OPE result $\frac{\Gamma}{|V_{cb}|^2}\times10^{13} = 5.7(9)$~GeV. For the ETM case, we quote the lattice result as $\frac{\Gamma}{|V_{cb}|^2}\times10^{13} = 0.987(60)$~GeV, while the OPE result is $\frac{\Gamma}{|V_{cb}|^2}\times10^{13} = 1.20(46)$. This shows that also after the $\boldsymbol{q}^2$ integration the lattice and OPE results remain compatible within the uncertainties.

In our work we also computed the lepton energy moments and the hadronic mass moments for which experimental results are already available. However, as in the case of the inclusive decay rate, we cannot yet compare the lattice results with expermient due to the use of unphysical masses in our lattice simulations. Instead, we can show the comparison between the lattice results and the OPE in the same fashion as previously done with the inclusive decay rate. The final results of the first lepton energy moment are $\langle E_l \rangle = 0.650(40)$~GeV for JLQCD and $\langle E_l \rangle = 0.491(15)$~GeV for ETMC which can be compared with the respective OPE predictions of $\langle E_l \rangle = 0.626(36)$~GeV and $ \langle E_l \rangle = 0.441(43)$~GeV respectively.
The results of the first hadronic mass moment are $\langle M^2_X \rangle = 3.75(31)$~GeV$^2$ for JLQCD and $\langle M^2_X \rangle = 3.62(14)$~GeV$^2$ at which corresponds the OPE predictions of $\langle M^2_X \rangle = 4.22(30)$~GeV$^2$ and $\langle M^2_X \rangle =4.32(56)$~GeV$^2$.
In general, for all the quantities computed in our study, we find a good agreement between the lattice data and the OPE curves expecially at low and moderate $q^2$. More details about the moments can be found in~\cite{Gambino:2022dvu}.

These findings provide an important and non-trivial test of the method used in this work, making us optimistic about the possibility of having soon a full lattice QCD computation of inclusive semileptonic decays, at a level of precision competitive with those from the OPE. Most importantly, the hope for the future is that full lattice QCD computations will allow to determine the inclusive value of $|V_{cb}|$ much more precisely than what is currently possible, and hopefully to clarify the origin of the persisting $3 \sigma$ tension between inclusive and exclusive determinations of this quantity. Moreover, a comparison between a complete lattice QCD calculation and the results from the OPE could allow one to test the quark-hadron duality on which the analytic method is based: this would be particularly important for inclusive $D$-meson decays, as the OPE converges more slowly in calculations of $c$ decays.

\section{Conclusions}
We presented our recent computation of inclusive decay rates from lattice QCD~\cite{Gambino:2022dvu}, which is based on the approach proposed in ref.~\cite{Gambino:2020crt}. As we discussed, our calculation is based on the numerical reconstruction of the integration kernel that regulates the integral over phase space of the structure functions contributing to the hadronic tensor. We applied two different methods for the kernel reconstruction, analyzing four-point Euclidean correlation functions obtained from two different ensembles of lattice QCD configurations (based on different discretization schemes for the gauge and quark fields, at $b$-quark masses lighter than in nature). The results that we obtained are consistent and can be successfully compared with analytic predictions from the OPE approach. This opens the path for systematic \emph{ab initio} non-perturbative studies of inclusive decays on the lattice, which hopefully will help solve the $V_{cb}$ puzzle.

\section*{Acknowledgements}
The numerical calculations of the JLQCD collaboration were performed on the SX-Aurora TSUBASA at the High Energy Accelerator Research Organization (KEK) under its Particle, Nuclear and Astrophysics Simulation Program, as well as on the Oakforest-PACS supercomputer operated by the Joint Center for Advanced High Performance Computing (JCAHPC). We thank the members of the JLQCD collaboration for sharing the computational framework and lattice data, and Takashi~Kaneko in particular for providing the numerical data for the exclusive decay form factors. The numerical simulations of the ETM collaboration were run on machines of the Consorzio Interuniversitario per il Calcolo Automatico dell'Italia Nord Orientale (CINECA) under the specific initiative INFN-LQCD123. A.S. thanks the organisers of the $51st$ ISMD conference for the possibility of presenting this work in the plenary session and the conference sponsors for granting financial support for his in person attendance.

\paragraph{Funding information}
The work of S.H. is supported in part by JSPS KAKENHI Grant Number JP26247043 and by the Post-K and Fugaku supercomputer project through the Joint Institute for Computational Fundamental Science (JICFuS). The work of P.G., S.M., F.S., S.S. is supported by the Italian Ministry of Research (MIUR) under grant PRIN 20172LNEEZ. This project has received funding from the Swiss National Science Foundation (SNF) under contract 200020\_204428.

\bibliography{inclusive_ISMD.bib}

\nolinenumbers

\end{document}